\documentclass[11pt]{article}

\usepackage[preprint]{acl}

\usepackage{times}
\usepackage{latexsym}

\usepackage[T1]{fontenc}

\usepackage[utf8]{inputenc}

\usepackage{microtype}

\usepackage{inconsolata}

\usepackage{graphicx}

\usepackage{dsfont}
\usepackage{xcolor}
\usepackage{subcaption}
\usepackage{float}
\usepackage{multirow}
\usepackage{array}
\usepackage{verbatim}
\usepackage{soul}
\usepackage{url}
\usepackage{amsmath}
\usepackage{amsthm}
\usepackage{amsfonts}
\usepackage{booktabs}
\usepackage{algorithm}
\usepackage{algorithmic}
\usepackage[switch]{lineno}
\usepackage{tcolorbox}
\usepackage{enumitem}

\usepackage{booktabs} 
\usepackage[table]{xcolor} 
\usepackage{hyperref}
\usepackage[percent]{overpic}
\usepackage{tikz}
\usepackage{array}
\title{CentaurTA Studio: A Self-Improving Human-Agent Collaboration System for Thematic Analysis}

\author{Lei Wang \\
  Temple University \\
  \texttt{tom.lei.wang@temple.edu} \\
  \And
  Min Huang \\
  Independent Researcher \\
  \texttt{minerstudy@gmail.com} \\
  \And
  Eduard Dragut \\
  Temple University \\
  \texttt{edragut@temple.com} \\
  }

\begin{document}
\maketitle
\begin{abstract}
Thematic analysis is difficult to scale: manual workflows are labor-intensive, while fully automated pipelines often lack controllability and transparent evaluation. We present \textbf{CentaurTA Studio}, a web-based system for self-improving human--agent collaboration in open coding and theme construction. The system integrates (1) a two-stage human feedback pipeline separating simulator drafting and expert validation, (2) persistent prompt optimization that distills validated feedback into reusable alignment principles, and (3) rubric-based evaluation with early stopping for process control.

Across three domains, CentaurTA achieves the strongest performance in both Open Coding and Theme Construction, reaching up to 92.12\% accuracy and consistently outperforming baseline systems. Agreement between the rubric-based LLM judge and human annotators reaches substantial reliability (average $\kappa = 0.68$). Ablation studies show that removing the feedback loop reduces performance from 90\% to 81\%, while eliminating the Critic or early stopping degrades accuracy or increases interaction cost. The full system reaches peak performance within 10 iterative rounds (about 25 minutes), demonstrating improved efficiency over expert-only refinement.
\end{abstract}

\section{Introduction}

Qualitative researchers increasingly experiment with large language models (LLMs) as analytic collaborators. Among qualitative methods, Thematic Analysis (TA) has emerged as a primary testing ground for such integration \cite{10.1109/WSESE66602.2025.00013}. TA relies on iterative cycles of coding, comparison, abstraction, and theme refinement \cite{braun2006using}. While flexible and widely adopted, these cycles demand sustained interpretive effort and careful documentation of analytic decisions to maintain rigor and transparency \cite{Rietz-20UIST-Cody,nowell2017thematic}. 

Existing LLM-assisted systems largely fall into two categories. Some focus on productivity gains in early analytic stages—for example, recommending candidate codes \cite{Gao-23CHI-CoAIcoder} or structuring collaborative coding pipelines \cite{Gao-24CHI-CollabCoder}. Others pursue higher levels of automation, including multi-agent frameworks that perform TA end-to-end without human participation \cite{Qiao-25WebConf-ThematicLM}. Although these approaches improve efficiency, they expose a central tension: automation may accelerate coding, yet qualitative inquiry fundamentally depends on human interpretation, contextual grounding, and reflexive judgment.

We introduce \textbf{CentaurTA Studio}\footnote{Software, prompts, sample data, and source code: \url{https://github.com/Zhuifeng414/CentaurTA_Studio/}.}\footnote{Demonstration video: \url{https://youtu.be/Zz7cxp5Nklk}.}, a deployable software system that operationalizes iterative, self-improving human--agent collaboration for thematic analysis \cite{mondal-etal-2024-dimsim,chen-etal-2024-iteralign,wang-etal-2024-e2cl,nayak2024long}. 

This demonstration paper presents:

\begin{itemize}
    \item \textbf{An end-to-end interactive system:} A web-based platform implementing structured Open Coding and Theme Construction through an Actor--Critic human--agent architecture with persistent principle learning.
    
    \item \textbf{Rubric-grounded process control:} A constraint-based evaluation module that supports expert assessment, automated scoring, item-level diagnostics, and early stopping.
    
    \item \textbf{Empirical validation:} Quantitative and human evaluation results showing that iterative expert feedback improves alignment, stability, and thematic coherence across domains.
\end{itemize}
\section{CentaurTA Studio Overview}
\label{sec:CentaurTA}

\begin{figure*}[t]
\centering
\includegraphics[width=0.7\textwidth]{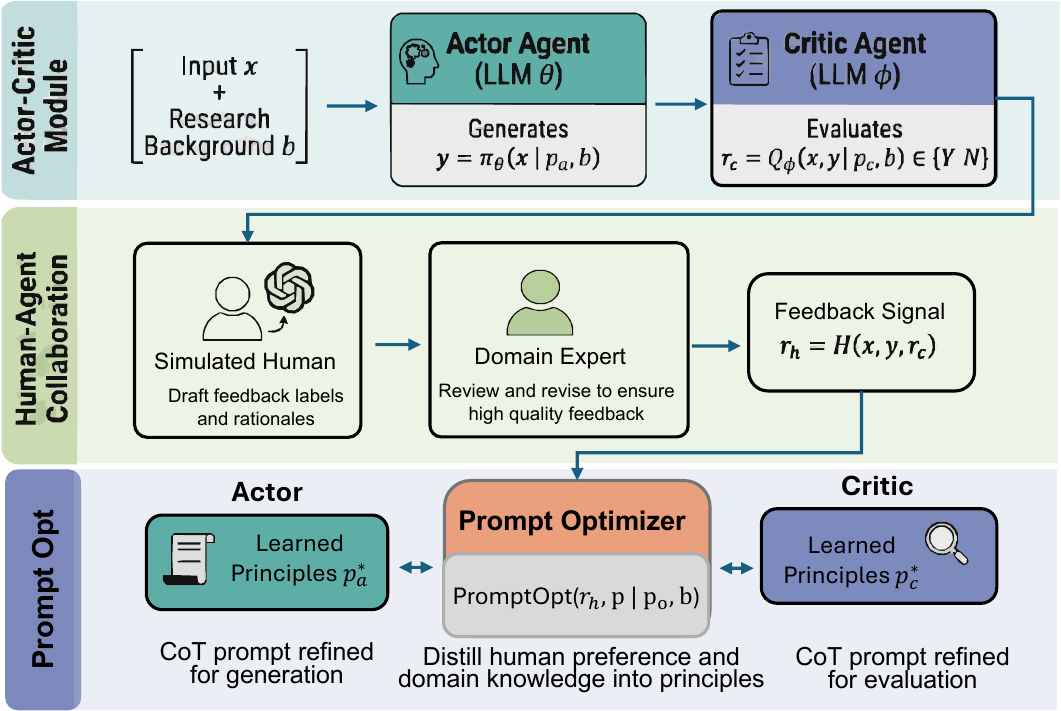}
\caption{The CentaurTA Studio framework integrates an Actor–Critic module, human–agent collaboration, and prompt optimization into a unified iterative loop. Through this closed-loop process, CentaurTA progressively aligns agents behavior with human subjective, knowledge, and preference, enabling iterative self-improving human–agent collaboration for qualitative analysis.}
\label{fig:AgentFramework}
\end{figure*}

CentaurTA Studio is an interactive human--agent workflow system for qualitative analysis.
Unlike a single static LLM invocation, the system organizes generation, evaluation, and expert feedback into a closed-loop process that progressively refines agent behavior.
Figure~\ref{fig:AgentFramework} presents the three-layer architecture:
(1) an Actor--Critic module,
(2) human--agent collaboration,
and (3) prompt optimization.

\subsection{Tasks and Output Schemas}
Let \( \mathcal{D} \) be the project corpus and \(b\) the study background. Each document \(d \in \mathcal{D}\) is mapped to an ordered sentence sequence \( \textit{sent}(d)=\{s_1,\ldots,s_n\} \). The system emits two artifact types.

\noindent \textbf{Open coding.} For each document segment, the system generates coding units
\( (code, quote, ref) \), where \(code\) is an inductive label, \(quote \subseteq d\) is evidence text, and \(ref\) points to supporting sentence IDs.

\noindent \textbf{Theme construction.} Using the accumulated code set \( \mathcal{C} \), the system builds higher-level units
\( (\textit{theme}, \textit{def}, \textit{codes}, \textit{rationale}) \), where \textit{codes} references supporting open codes and \textit{rationale} records why they are grouped.

\subsection{Actor--Critic Module}

Given task input $x$ and research background $b$, the system invokes two role-specific agents.

\paragraph{Actor Agent.}
The Actor $\pi_\theta$ generates analytical outputs:
$
y = \pi_\theta(x \mid p_a, b),
$
where $p_a$ is the generation prompt.

\paragraph{Critic Agent.}
The Critic $Q_\phi$ evaluates the generated output:
$
r_c = Q_\phi(x, y \mid p_c, b),
$
where $p_c$ defines evaluation criteria.
The Critic produces structured feedback indicating whether rubric constraints are satisfied.

\subsection{Two-Stage Human Feedback}

CentaurTA integrates human supervision through a two-stage design.

\paragraph{Stage 1: Simulated Human.}
A simulated reviewer produces draft feedback labels and rationales. 
This stage reduces repetitive workload and highlights uncertain cases.

\paragraph{Stage 2: Domain Expert.}
Domain experts review and revise the draft feedback in real time.
The finalized supervision signal is denoted as:
$
r_h = H(x, y, r_c),
$
where $H(\cdot)$ represents expert validation.
Only expert-confirmed feedback is used for refinement, ensuring that domain interpretation remains authoritative.

\subsection{Prompt Optimization}

Instead of updating model weights, CentaurTA refines prompts through a Prompt Optimizer.
Validated feedback is distilled into updated role-specific principles:
$
p_a^{*} = \mathrm{PromptOpt}(r_h, p_a),
$,
$
p_c^{*} = \mathrm{PromptOpt}(r_h, p_c).
$
The updated prompts are versioned and applied to subsequent batches, forming an iterative improvement loop.

\begin{figure*}[t]
\centering
\includegraphics[width=0.9\textwidth]{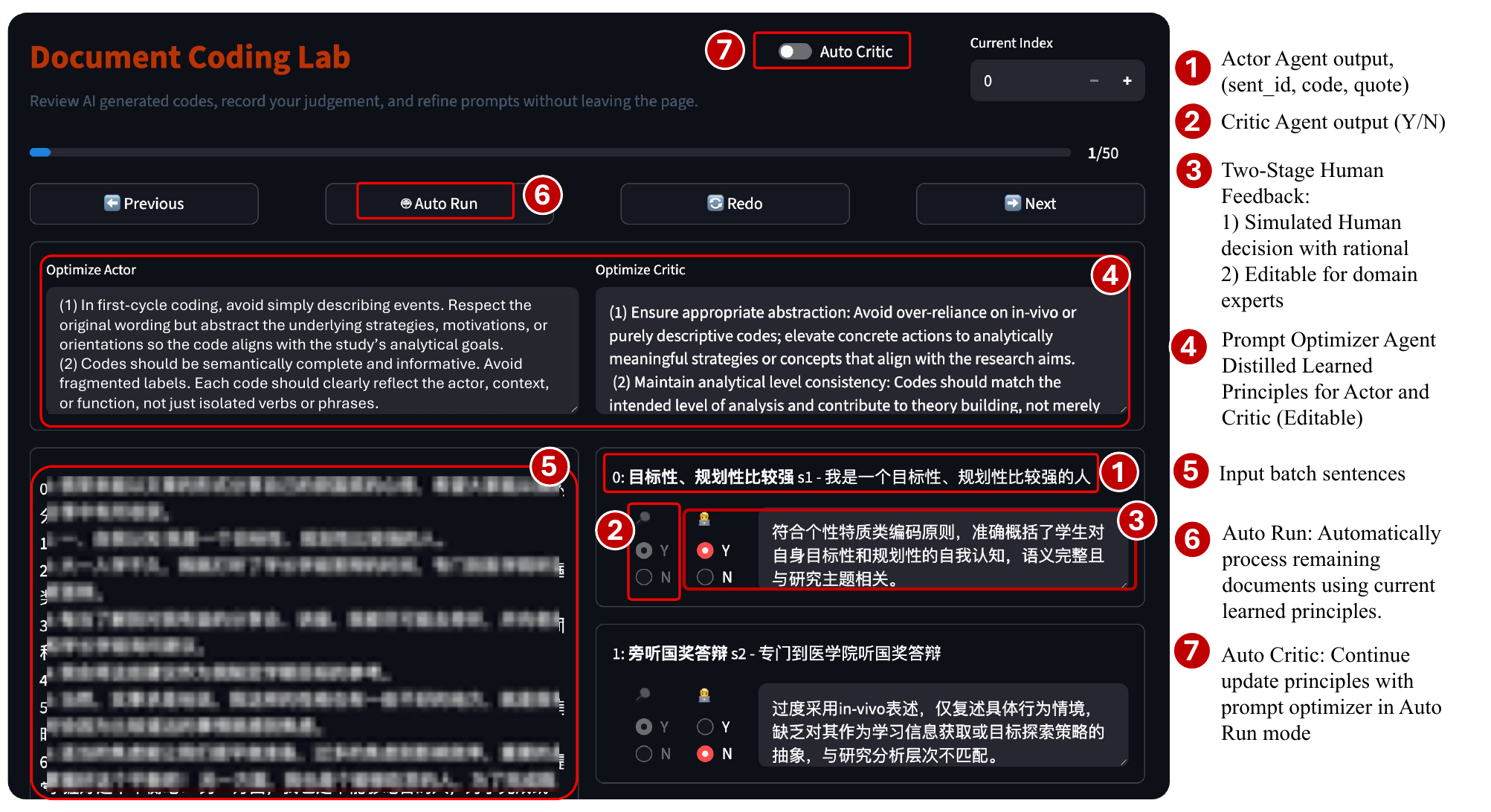}
\caption{Interface of the Open Coding Lab. The system displays Actor-generated structured codes, independent Critic judgments, two-stage human feedback, and an editable Prompt Optimizer. Auto Run and Auto Critic enable scalable iterative refinement.}
\label{fig:OpenCoding}
\end{figure*}

\begin{figure*}[t]
\centering
\includegraphics[width=0.8\textwidth]{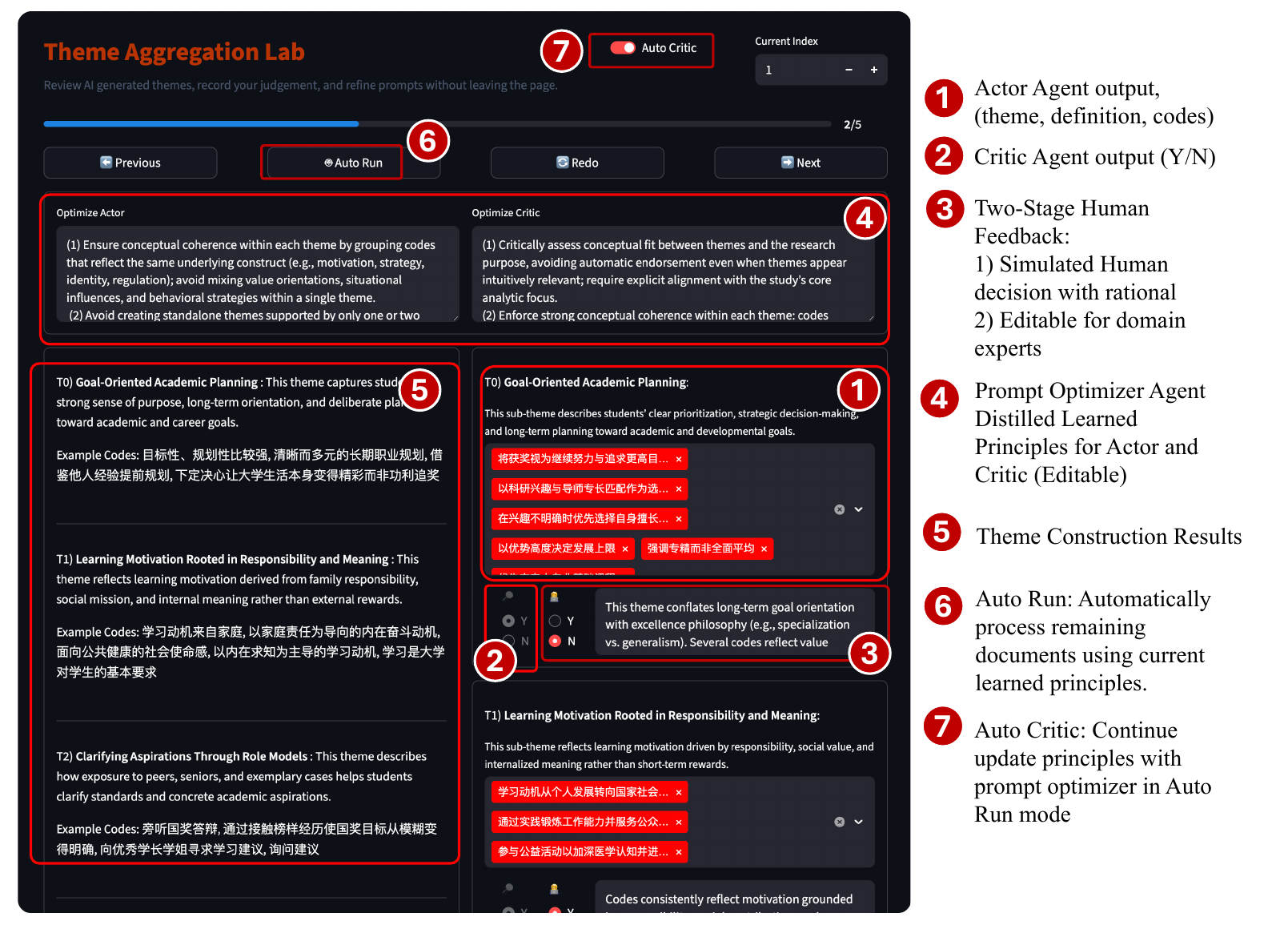}
\caption{Interface of the Theme Aggregation Lab. Themes are constructed from validated codes with explicit definitions and references. Actor–Critic separation, expert-editable feedback, and automated prompt updating support traceable abstraction from codes to themes.}
\label{fig:ThemeConstruction}
\end{figure*}

\section{Software Implementation}

CentaurTA Studio is an interactive web-based system that operationalizes iterative human--agent collaboration for qualitative analysis. The interface is organized into two structured workspaces: the Document Coding Lab (Figure~\ref{fig:OpenCoding}) and the Theme Aggregation Lab (Figure~\ref{fig:ThemeConstruction}). Each lab enforces a fixed analytical schema and visualizes the generation--evaluation loop explicitly.

\noindent\textbf{Open Coding}
As illustrated in Figure~\ref{fig:OpenCoding}, the system processes input sentences in batches (5). The \emph{Actor Agent} generates structured outputs in the form $(sent\_id, code, quote)$ (1), where \textit{quote} must be a span grounded in the original text. The \emph{Critic Agent} independently evaluates each code with a binary judgment (2). A two-stage human feedback module (3) presents simulated decisions with rationales that remain editable by domain experts. A \emph{Prompt Optimizer} (4) distills expert-validated feedback into explicit principles for Actor and Critic Agents that are transparently displayed and editable. The \textit{Auto Run} and \textit{Auto Critic} modes (6,7) enable large-scale processing while continuously updating prompt principles.

\noindent\textbf{Theme Construction}
Figure~\ref{fig:ThemeConstruction} shows the second-stage abstraction interface. Here, the Actor produces structured outputs $(theme, definition, codes)$ (1), and the Critic evaluates conceptual coherence (2). Each theme must reference supporting codes and provide justification. Similar to open coding, expert-editable feedback (3) and a Prompt Optimizer (4) refine theme-level principles. The interface displays full theme construction results (5) while supporting iterative auto-processing (6,7).

Actor generation and Critic evaluation are strictly separated at the interface and prompt levels in both labs. Only expert-confirmed judgments are retained and distilled into updated prompt principles. Through structured outputs, role separation, and expert-validated prompt refinement, CentaurTA implements a transparent, iterative human--agent workflow without modifying model parameters.

\begin{table}[ht]
\centering
\resizebox{0.48\textwidth}{!}{
\begin{tabular}{lclccc}
\toprule
\textbf{Dataset}  &  \textbf{\# Doc} & \textbf{Lang} & \textbf{\# w} & \textbf{\# sent} & \textbf{Avg \# w} \\
\hline
USRS  & 12 & Ch & 25,718 & 471 & 2,143 \\ 
ASP  & 15 & Eng & 12,678 & 651 & 845 \\
Dreaddit  & 214 & Eng & 19,358 & 822 & 98 \\
\bottomrule
\end{tabular}
}
\caption{Data statistics }
\label{tab:sta_dataset}
\end{table}

\section{Experimental Design}
\label{sec:exp_setup}

\subsection{Datasets}
We use 3 domain-specific datasets. 
\textbf{USRS}: this dataset focus on self-regulated learning of Chinese undergraduates, which consists of 12 personal Reflection texts written in Chinese by National Scholarship recipients over the past two years from University H \cite{Min-PhDThesis2024-Beyond}.
\textbf{ASP}: this dataset explores the challenges and assistive technology for autistic job seekers \cite{GARRISON2025105155}.
\textbf{Dreaddit}: a public English-language corpus of short, multi-domain social media texts for stress detection \cite{Turcan-LOUHI19-Dreaddit}.
USRS and ASP have received IRB approval. All datasets were anonymized prior to analysis.
Table~\ref{tab:sta_dataset} summarizes the dataset statistics used in our experiments.

\subsection{Baselines Selection}

We evaluate the performance of our CentaurTA framework against 2 baselines:
1) MindCoder \cite{Gao-25arXiv-MindCoder}, an LLM-powered workflow for qualitative analysis, while enabling humans to conduct meaningful interpretation.
2) Atlas.ti \cite{ATLAS-2025-AIcoding}, a commercial platform for AI Coding.


\subsection{Rubric-based Evaluation}

We evaluate system outputs using a structured 30-constraint rubric library derived from qualitative research methodology. 
For open coding, each \{code, quote\} pair is assessed using 18 constraints (6 general rules, 7 format/linguistic constraints, and 5 domain-specific requirements). 
For theme construction, each theme is evaluated with 12 constraints (4 general theme rules, 4 format constraints, and 4 domain-specific requirements). 
An LLM-based judge verifies whether each constraint is satisfied, and the final rubric score is computed as the average constraint satisfaction rate. 
This fine-grained evaluation provides interpretable diagnostics and supports principled early stopping during self-improvement.

\subsection{Expert Participation}

We recruited six participants through institutional mailing lists and professional networks. 
Two self-identified as experts in thematic analysis, and four as intermediate practitioners. 
Participants were divided into two disjoint groups to prevent evaluation leakage. 
Three annotators participated in the CentaurTA human-in-the-loop feedback process, while the remaining three formed an independent expert panel responsible for curating the verifiable rubric library used for rubric-based evaluation. 
The rubric panel did not participate in the interactive feedback process, and annotators in the human-in-the-loop condition were not exposed to the rubric library. 
Rubric curation achieved substantial inter-rater agreement ($\kappa = 0.78$), indicating reliable expert consensus.

\begin{table}[t]
\centering
\small
\begin{tabular}{lcccccc}
\toprule
PID & TA Exp. & Years  & Position & HF & RC \\
\midrule
P1 & Expert & 10  & Prof. & Y & N \\
P2 & Int. & 4  & PhD & Y & N \\
P3 & Int. & 3  & PhD & Y & N \\
P4 & Expert & 7 & Prof. & N & Y \\
P5 & Int. & 5  & PhD & N & Y \\
P6 & Int. & 4 & PhD & N & Y \\
\bottomrule
\end{tabular}
\caption{Expert roles and participation. Int. denotes intermediate practitioner; HF indicates human-feedback participation; RC indicates rubric curation.}
\label{tab:experts}
\end{table}

\section{Results}

\subsection{Effectiveness of Human–Agent Collaboration}
\label{sec:human_agent_collab}
\begin{figure*}[t]
\centering
\includegraphics[width=0.7\textwidth]{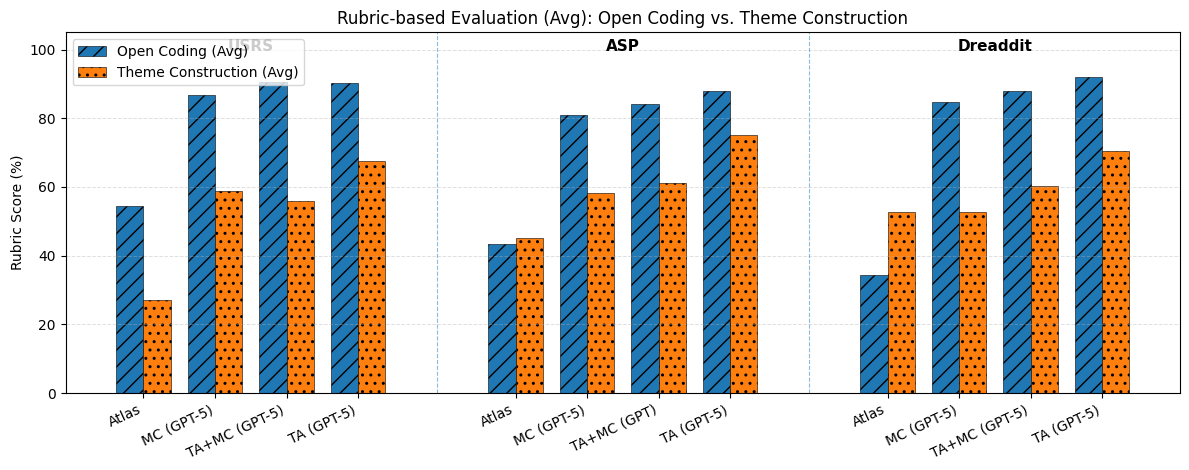}
\caption{Rubric-based evaluation (average scores) comparing Open Coding and Theme Construction across three datasets (USRS, ASP, Dreaddit).}
\label{fig:MainRes}
\end{figure*}

Figure \ref{fig:MainRes} compares MindCoder (MC), TA+MC (learned principles of TA are applied to MC), and TA on Open Coding and Theme Construction across three domains. TA consistently achieves the strongest performance, improving over MC in Open Coding on USRS (90.21 vs. 86.81), ASP (87.78 vs. 81.05), and Dreaddit (92.12 vs. 84.62). For Theme Construction, the gains are more pronounced. Importantly, TA+MC also improves over MC alone, for example in Open Coding on USRS (90.40 vs. 86.81) and Dreaddit (88.05 vs. 84.62). These results indicate that the learned alignment principles generalize across platforms, improving both code-level accuracy and higher-level thematic abstraction.

\subsection{Agreement Between LLM Judge and Human Annotators}
To further validate the reliability of rubric-based evaluation, we conducted an additional agreement study in the USRS domain. A domain expert independently labeled 300 open-coding outputs, and we compared these annotations with the decisions produced by the rubric-based LLM-as-Judge on the same items. Agreement was assessed using Cohen’s $\kappa$ and constraint-level accuracy.

As reported in Table~\ref{tab:judge_agreement}, the LLM judge achieves an average rubric accuracy of 90\%, closely matching a human expert’s average accuracy of 89\%. The overall inter-annotator agreement between the human expert and the LLM judge reaches $\kappa = 0.68$, indicating substantial agreement. Agreement is highest for general constraints ($\kappa = 0.71$) and moderately lower for domain-specific constraints ($\kappa = 0.64$), which require more nuanced interpretation.

\begin{table}[t]
\centering
\footnotesize
\setlength{\tabcolsep}{3pt}
\resizebox{0.95\columnwidth}{!}{
\begin{tabular}{lccc}
\toprule
\textbf{Rubric Type} & \textbf{Rubric Acc} & \textbf{Human Acc} & \textbf{Human--LLM $\kappa$} \\
\midrule
General rules       & 92 & 93 & 0.71 \\
Format / linguistic & 94 & 91 & 0.68 \\
Domain-specific     & 84 & 82 & 0.64 \\
\midrule
\textbf{Avg}        & \textbf{90} & \textbf{89} & \textbf{0.68} \\
\bottomrule
\end{tabular}
}
\caption{Agreement between rubric-based LLM judge and human annotators.}
\label{tab:judge_agreement}
\end{table}

\subsection{Ablation and Efficiency}

We conduct controlled experiments on the Open Coding task in the USRS domain, using a batch size of 10 sentences and a maximum of 25 iterations. The experiments aim to (1) separate the contributions of simulated and domain-expert feedback, and (2) analyze the impact of major system components.

\paragraph{Feedback Separation.}

\begin{table}[t]
\centering
\footnotesize
\setlength{\tabcolsep}{3pt}
\resizebox{0.95\columnwidth}{!}{
\begin{tabular}{lccc}
\toprule
\textbf{Exp} & \textbf{Acc (\%)} & \textbf{Time (m)} & \textbf{Rnd} \\
\midrule
\textbf{CentaurTA} & 90 & 25 & 10 \\
w/o human feedback & 81 & 3  & -- \\
w/o simulated (expert-only) & 90 & 42 & 10 \\
w/o domain expert (sim-only) & 85 & 7 & 5 \\
\bottomrule
\end{tabular}
}
\caption{Feedback separation analysis.}
\label{tab:feedback_ablation}
\end{table}
Removing the feedback loop reduces performance to 81\%, confirming that gains arise from iterative alignment rather than static prompting. The expert-only setting achieves comparable accuracy (90\%) but at substantially higher cost (42 minutes), while the simulator-only setting is more efficient (7 minutes) but less accurate (85\%). These results show that alignment gains stem from the two-stage human feedback mechanism, not from the simulator alone.

\paragraph{Component Ablations.}

\begin{table}[t]
\centering
\footnotesize
\setlength{\tabcolsep}{3pt}
\resizebox{0.95\columnwidth}{!}{
\begin{tabular}{lccc}
\toprule
\textbf{Exp} & \textbf{Acc (\%)} & \textbf{Time (m)} & \textbf{Rnd} \\
\midrule
\textbf{CentaurTA}        & 90 & 25  & 10 \\
w/o prompt opt.           & 92 & 180 & -- \\
w/o critic agent          & 87 & 23  & 12 \\
w/o rubric early stop     & 88 & 75  & 25 \\
\bottomrule
\end{tabular}
}
\caption{Component ablation analysis.}
\label{tab:component_ablation}
\end{table}

Prompt optimization primarily improves efficiency: without it, total time increases substantially (180 minutes). The self-improving mechanism distills learned principles into the Actor and Critic prompts, reducing repeated errors and alleviating redundant expert corrections. Removing the Critic reduces performance to 87\%, demonstrating that the actor--critic decomposition provides structured evaluation signals that stabilize learning. Rubric-based early stopping mitigates overfitting and reduces unnecessary interaction rounds.
\section{Related Work}

\paragraph{LLM-Assisted Thematic Analysis.}
Recent work integrating large language models into thematic analysis (TA) follows two main directions. One emphasizes \emph{interactive analyst support}, embedding LLM suggestions into collaborative coding tools such as CoAIcoder, CollabCoder, and MindCoder \cite{Gao-23CHI-CoAIcoder,Gao-24CHI-CollabCoder,Gao-25arXiv-MindCoder}. These systems improve usability and drafting efficiency but generally treat human feedback as session-level correction rather than persistent supervision.
The other direction focuses on \emph{multi-agent automation}, where coordinated agents generate codes and themes at scale with limited human involvement \cite{Qiao-25WebConf-ThematicLM,Xu-25arXiv-TAMA,Yi-25arXiv-AutoTA}. While scalable, such approaches raise concerns about interpretive validity and alignment with qualitative research practices.

\paragraph{Self-Improving and Alignment-Driven LLM Agents.}
Beyond qualitative analysis, a growing body of work investigates mechanisms for iterative self-improvement in LLM-based agents. Recent studies propose frameworks in which model behavior is refined through structured feedback, prompt optimization, or preference-based alignment \cite{mondal-etal-2024-dimsim,chen-etal-2024-iteralign,wang-etal-2024-e2cl,nayak2024long}. Related discussions in human-in-the-loop research further argue that effective collaboration requires bidirectional adaptation rather than static assistance \cite{shen2025bidirectional}. While such self-improving paradigms have shown promise in reasoning and alignment tasks, they have not yet been systematically applied to thematic analysis workflows. CentaurTA bridges this gap by integrating rubric-guided expert supervision directly into an iterative Actor--Critic loop tailored to qualitative coding and theme construction.

\section{Conclusions}
Overall, the results show that CentaurTA consistently improves performance through structured human–agent collaboration. TA outperforms both MC and TA+MC across tasks, while TA+MC’s gains over MC demonstrate that learned alignment principles generalize across domains. Agreement analysis supports the reliability of rubric-based evaluation, and ablations confirm that improvements stem from the two-stage feedback mechanism and system design rather than prompt tuning alone. Together, these findings validate CentaurTA as an effective and scalable alignment framework.

As a live system, CentaurTA Studio supports batch execution, interactive correction, principle updates, and exportable analysis artifacts. The Open Coding Lab and Theme Aggregation Lab separate low-level coding from higher-level abstraction, making the analytical process transparent and modular. 

\section{Limitations}

CentaurTA Studio still requires expert validation for principle updates, as domain interpretations cannot be fully automated. The effectiveness of rubric-based evaluation depends on the quality and completeness of the rubric design itself.

\bibliography{bibs/Background, bibs/Method}

\newpage
\appendix

\section*{Ethical Considerations}
This work studies LLM-assisted thematic analysis of qualitative text and proposes a human--agent collaboration framework that preserves human interpretive control. The study does not involve intervention, deception, or collection of personal identifiers beyond expert-in-the-loop feedback for evaluating model outputs.

We evaluate CentaurTA on three datasets spanning education, assistive employment, and social media. Two datasets (USRS and ASP) received prior IRB approval in their original studies, and all datasets were anonymized before analysis. We follow dataset licenses and usage terms and do not release new raw personal data.

Because qualitative analysis is interpretive, automation can introduce bias or over-abstraction. To mitigate this, CentaurTA requires evidence grounding for codes, applies rubric-based methodological constraints, and keeps human experts as final decision-makers through explicit override controls.

Finally, we acknowledge environmental and computational costs associated with LLM usage. The framework adapts behavior through prompt updates rather than parameter finetuning, reducing compute overhead during iterative refinement.

\end{document}